\documentclass[pra,aps,showpacs,amsmath,amssymb]{revtex4}
\usepackage{color}
\usepackage{amsmath}
\usepackage{graphicx}

\makeatletter
\@ifundefined{textcolor}{}
{%
 \definecolor{BLACK}{gray}{0}
 \definecolor{WHITE}{gray}{1}
 \definecolor{RED}{rgb}{1,0,0}
 \definecolor{GREEN}{rgb}{0,1,0}
 \definecolor{BLUE}{rgb}{0,0,1}
 \definecolor{CYAN}{cmyk}{1,0,0,0}
 \definecolor{MAGENTA}{cmyk}{0,1,0,0}
 \definecolor{YELLOW}{cmyk}{0,0,1,0}
 }

\@ifundefined{definecolor}
 {\usepackage{color}}{}
\@ifundefined{definecolor}
 {\@ifundefined{definecolor}
 {\usepackage{color}}{}
}{}
\begin{document}

\title{The quantum cryptographic switch}

\author{ Srinatha Narayanaswamy and Omkar Srikrishna}

\affiliation{Poornaprajna Institute of Scientific Research, Sadashivnagar, Bengaluru-
560080, India.}

\author{R. Srikanth}

\affiliation{Poornaprajna Institute of Scientific Research, Sadashivnagar, Bengaluru-
560080, India}

\affiliation{Raman Research Institute, Sadashivnagar, Bengaluru- 560060, India.}

\author{Subhashish Banerjee}

\affiliation{ Indian Institute of Technology Rajasthan, Jodhpur-342011, India.}

\author{Anirban Pathak}

\affiliation{Jaypee Institute of Information Technology, A-10, Sector-62, Noida,
Uttar Pradesh-201307, India.}
\begin{abstract}
We illustrate using a quantum  system the principle of a cryptographic
switch, in which a third party (Charlie) can control to a continuously
varying degree the amount  of information the receiver (Bob) receives,
after  the sender (Alice)  has sent  her information.  Suppose Charlie
transmits a Bell  state to Alice and Bob.  Alice  uses dense coding to
transmit two bits to Bob.  Only if the 2-bit information corresponding
to choice  of Bell state is made  available by Charlie to  Bob can the
latter  recover Alice's  information.  By  varying the  information he
gives, Charlie can continuously vary the information recovered by Bob.
The performance of the  protocol subjected to the squeezed generalized
amplitude damping channel  is considered. We also present  a number of
practical situations where a cryptographic switch would be of use.
\end{abstract}


\maketitle

\section{Introduction}

In 1984, Bennett and Brassard first introduced a protocol \cite{bb-84}
for quantum key distribution (QKD). In QKD two remote legitimate users
(Alice and  Bob) can establish an unconditionally  secured key through
the transmission of  qubits. Since the pioneering work  of Bennett and
Brassard several protocols for different cryptographic tasks have been
proposed.  While most  of the  initial works  on  quantum cryptography
\cite{bb-84,Ekert,B-92}  were   concentrated  around  QKD,  eventually
quantum-states  were  applied  to other  `post-coldwar'  cryptographic
tasks.  For  example,  in  1999,  Hillery  \cite{Hillery}  proposed  a
protocol for quantum  secret sharing (QSS). In the  same year, Shimizu
and  Imoto \cite{Imoto}  proposed  a protocol  direct secured  quantum
communication   using   entangled    photon   pairs.   Protocols   for
deterministic secured quantum communication (DSQC) were later proposed
\cite{Ping-pong,vaidman-goldenberg,LM-05},  in which the  receiver can
read out  the secret message only  after the transmission  of at least
one bit of  additional classical information for each  qubit. A set of
protocols  exist   which  does  not  require   exchange  of  classical
information. Such protocols are generally referred to as protocols for
``quantum secure direct communication'' (QSDC) \cite{reiview}.

DSQC and  QSDC protocols are reducible to  secure QKD protocols
in  the  sense that  the  former equipped  with  a  source of  quantum
randomness, yield  the latter.  A conventional  QKD protocol generates
the  unconditionally  secured  key  by  quantum means  but  then  uses
classical  cryptographic  resources to  encode  the  message. No  such
classical means are required in  DSQC and QSDC.  In recent past, these
facts have encouraged several groups  to study DSQC and QSDC protocols
in detail, see for example \cite{reiview} (and references therein).

Since the works of Shimizu and Imoto \cite{Imoto}, and Hillery {\em et
  al.} \cite{Hillery}, several protocols of DSQC and QSS are proposed.
The  unconditional security  of  the protocols  is  achieved by  using
different quantum  resources. For example, DSQC  schemes are proposed,
a)     with     and      without     maximally     entangled     state
\cite{dsqcqithout-max-entanglement} (and references therein), b) using
teleportation \cite{dsqcwithteleporta},  c) using entanglement swapping
\cite{entanglement  swapping},  d)  using  rearrangement  of  order  of
particles, etc.  We are  specifically interested in the DSQC protocols
based on rearrangement of orders  of particles.  To be precise here we
aim to provide  a protocol of DSQC for  a specific task/problem, which
may  be  visualized as  follows:  Charlie,  Alice  and Bob  are  three
employees of  a company.  Charlie is  the Director of  the company and
Alice is a  \textit{custodian} of all the files  of that company.  Now
Alice wishes to securely share some of the information with Bob, say a
file, which may be considered as  a string of classical bits.  For the
present work, we have chosen it to be classical information, though it
can be easily generalized to quantum information.  But to do so, Alice
requires permission from Charlie, who, though permitting  Alice to do
so, has  imposed a restriction  that Bob can  read the file  only when
Charlie  permits  him  to  do  so.  This is  a  problem  of  practical
importance because Alice  may transmit her quantum-encoded information
to Bob in advance (say,  because the quantum channel is available only
then), but  Charlie may wish to  give his permission at  a later time,
determined by, say, certain administrative constraints.

Here  we propose  a  protocol of  DSQC  to solve  the above  mentioned
problem    by   using    rearrangement   of    order    of   particles
\cite{reordering1,the:C.-W.-Tsai}        and        dense       coding
\cite{denscode}.  Since  a  secret  is  shared, the  protocol  may  be
considered  as  a  protocol  of  secret  sharing.  This  can  also  be
visualized as  an application of \textit{controlled  dense coding}, in
which  the controller is  Charlie, who  determines how  much classical
information is  delivered to Bob after  Alice sends him  all her dense
coding qubits. Since  the problem discussed here is  of much practical
importance,  we  made the  protocol  more  realistic  and relevant  by
considering the  channel to be noisy. By doing  so we generalize the
idea  of secret  sharing  through noisy  quantum  channel proposed  in
\cite{puncudge}.

The remaining part of the paper is organized as follows: In the next
section we have described a protocol that realizes a cryptographic
switch by using a pure entangled state (Bell state) in an ideal scenario.
In this Section we put some restriction on the channel between Alice
and Bob. In Section III, the effect of noise on the protocol is discussed
with specific attention to the squeezed generalized amplitude damping
channel. In Section IV we have modified the protocol for a more general
case where no restriction is imposed on the channel between Alice
and Bob. Finally, Section V is dedicated for conclusions and discussions.

\section{Cryptographic Switch with pure state}

Let us first propose a solution  for the above mentioned problem in an
ideal scenario: Here  Alice is semi-honest and Bob  can receive qubits
from Alice but  cannot send a qubit to Alice  (for example, because he
lacks a requisite license). Thus  the channel between Alice and Bob is
one-way (a  restriction that  we will relax  in Section IV).  To begin
with, we also  assume that the channel is  noiseless.  By semi-honest,
we mean that Alice strictly follows the protocol and does not disclose
the contents of the file to Bob before obtaining Charlie's permission.
In other  words since Alice  is honest to  the protocol, she  does not
create  a  quantum  channel  between  herself  and  Bob  to  send  the
information  through   that.   But   after  obtaining  her   share  of
entanglement  from Charlie  she  may try  to  help Bob  to bypass  the
control  of  Charlie and  to  obtain  the  information before  Charlie
permits Bob to do so. We do  not require Bob to be semi-honest as even
a  dishonest Bob  cannot obtain  any  information as  long as  Alice
respects  the protocol.  Here it  would be  apt to  note that  in the
context of  secure multiparty  communication, the semi-honest  and the
malicious models  are generally  used as secutiy  models. Here  we are
interested   about  the  semi-honest   model,  in   which  semi-honest
participants (Alice  in our case)  are assumed to follow  the protocol
(and are  thus honest)  but want  to learn any  other's secret  (in our
case, Alice  and Bob want to  cooperate to learn, which  Bell state is
prepared by Charlie;  thus they are curious).  In a semi-honest model,
a protocol  is considered secure against a  collusion of participants
(Alice  and Bob  in our  case) if  by accumulating  their  data, these
participants cannot gain more information than what  they can 
from    the    input  and    output   of    the    protocol    alone
\cite{brassard-semi-honest}.   In  contrast  to  this in  a  malicious
model, participants can deviate  from the orginial protocol (say Alice
and Bob  can use  LM05 protocol to  share the information  without any
involvement  of  Charlie). The  protocols  proposed  here and  similar
protocols are not secured under a malicious model.

The use of  the semi-honest model is well-jusitified  in classical and
quantum  communication  \cite{semi-honest  quantum}.   In  particular,
semi-honest   models  are   very  often   used  in   modern  classical
cryptography,    specially   in   the    context   of    data   mining
\cite{datamining-semi-honest}.  In  the recent  past, the notion  of a
semi-honest user  is also adopted  in the field of  multi-party secure
quantum  communication  and  different secured  quantum  communication
protocols          have          already         been          proposed
\cite{brassard-semi-honest,semi-honest  quantum},  where  one or  more
participants are  considered as semi-honest.   To begin with,  we will
first propose a  protocol for designing of cryptographic  switch in an
ideal scienario where  the channel between Alice and  Bob is noiseless
and one-sided.  The  one sidedness of the the  channel will essentially
exclude the possible act of collusion between Alice and Bob. In Section IV,
the restriction of one-sidedness of  the channel will be removed and 
then  the security  against  a collusion  of  Alice and  Bob would  be
obtained  by  rearrangement of  order  of  particles.   In this  ideal
situation our protocol works as follows:

\begin{enumerate}
\item After receiving Alice's request, Charlie prepares $n$ Bell states
(not all the same) and sends first qubits of all the Bell states to
Alice and the second qubits to Bob. But Charlie does not disclose
which Bell state he has prepared. 
\item After receiving the qubits from Charlie, Alice understands that she
has been permitted to send the information to Bob. 
\item Alice uses dense coding to encode two bits of classical information
on each qubit and transmits her qubits to Bob. 
\item When Charlie plans to allow Bob to know the secret information communicated
to him, he discloses the Bell state which he had prepared. 
\item Since the initial Bell state is known, Bob now measures his qubits
in Bell basis and obtains the information encoded by Alice. 
\end{enumerate}
Bob can perform Step 5 (i.e.,  measurement in Bell basis) before Step 4
but  he  will  not  obtain  any  meaningful  information  without  the
knowledge of the initial state.  A  Bell measurement can reveal  the Bell
state  prepared by Charlie  if both  the qubits  are in  possession of
Alice or Bob. Now since we have assumed that the channel between Alice
and Bob is one-way, therefore, Bob cannot send his qubits to Alice for
Bell  measurement.  Alice  cannot send  her qubits  to Bob  as  she is
assumed  to   be  a  semi-honest   party  who  strictly   follows  the
protocol. Her  semi-honesty is motivated  by the fact that,  while she
may wish to potentially cheat  Charlie, she wants her communication to
Bob to be secure both in the sense of being protected from the rest of
the world (in  the usual QKD sense) as well as being undetected
by Charlie.

Here we are considering  a three-party quantum  secret sharing scenario,  in which
Charlie prepares a Bell state, of which he transmits one half to Alice
and  the other half to  Bob.  From  this viewpoint,  the latter  receives an
ensemble of Bell-states:
\begin{equation}
\rho_{AB}\equiv\sum_{j,k}a_{j,k}|B_{j,k}\rangle\langle
B_{j,k}|,\label{eq:bell}\end{equation} $j$ denotes  the parity bit and
$k$                   the                  phase                  bit:
$|B_{0,0/1}\rangle\equiv\frac{1}{2}(|00\rangle\pm|11\rangle)$       and
$|B_{1,0/1}\rangle=\frac{1}{2}(|01\rangle\pm|10\rangle)$.         Alice
encodes two  bits on her qubit  using the four Pauli  operators of the
superdense coding protocol \cite{denscode}, and sends the qubit to Bob
via a possibly insecure channel.

In the noiseless  case, Bob measures the two  qubits in his possession
to obtain the state that corresponds to Alice's encoding. However, Bob
can  decode the  full  information  only if  Charlie  shares the  full
classical  \textit{key} information  $c$ that  would make  the initial
entangled state pure. More generally (as detailed below), Bob recovers
Alice's  transmitted bits  depending on  the key  information obtained
from Charlie. Thus Charlie acts as a \textit{cryptographic switch} who
can   determine  the  level   of  information   Alice  sends   to  Bob
\textit{after}    the   full   transmission    of   her    qubit.   In
Ref.  \cite{puncudge},  a related  protocol  is  considered, in  which
Charlie prepares a GHZ state, from which he sends a qubit to Alice and
one to  Bob. After Alice's  encoding and transmission to  Bob, Charlie
measures his qubit in the  $X$ basis, thereby collapsing the Alice-Bob
state   to  one   of  the   two  Bell   states   $|B_{0,0}\rangle$  or
$|B_{0,1}\rangle$.  Only  when  Charlie   reveals  his  1  bit  of  key
information does Alice's dense coding protocol succeed.

Note that Alice's and Charlie's operations commute, so that we might
as well consider that Charlie prepares one of these two Bell states,
transmitting them to Alice and Bob. Thus, our protocol generalizes
that of Ref. \cite{puncudge}, in which all four Bell states, instead
of the above two alone, are considered. The key information in our
case is thus 2 bits. We can thus consider a family of protocols in
which the key information $c$ varies continuously as $0\le c\le c_{{\rm max}}=2$.
Our protocol is characterized by $c=c_{{\rm max}}$ and that of Ref.
\cite{puncudge} by $c=1$.

Suppose Bob's  measurement yields the state  $|\Phi^{+}\rangle$. If he
comes   to    know   that   Charlie   had    transmitted   the   state
$|\Phi^{-}\rangle$, then  he knows Alice must have  encoded $Z$, since
$|\Phi^{+}\rangle=(Z\otimes I)|\Phi^{-}\rangle$.  More generally, this
idea    can   be    compactly    represented   as:    \begin{equation}
  B_{j,k}\stackrel{P_{ab}}{\longrightarrow}B_{j\oplus         a,k\oplus
    b},\hspace{1cm}(j,k,a,b=0,1)\label{eq:belltrans}\end{equation}
where  the l.h.s  is state  $B_{j,k}$  prepared by  Charlie, which  is
transformed  under Alice's action  $P_{ab}$ to  the state  received by
Bob, given in the r.h.s. The sign $\oplus$ denotes mod 2 addition, and
$P_{ab}$ ($a,b=0,1$) represents, sequentially, the Pauli $I$, $Z$, $X$
and $Y$ operators.

The problem can be treated as a communication situation in which Alice
is signaling Bob by means of Bell states. Then the maximum information
Bob can extract from the pair of qubits is the Holevo quantity \cite{nc00}.
In our protocol, where $c=2$, Bob can extract Alice's 2 bits of information.

\begin{figure}
\includegraphics{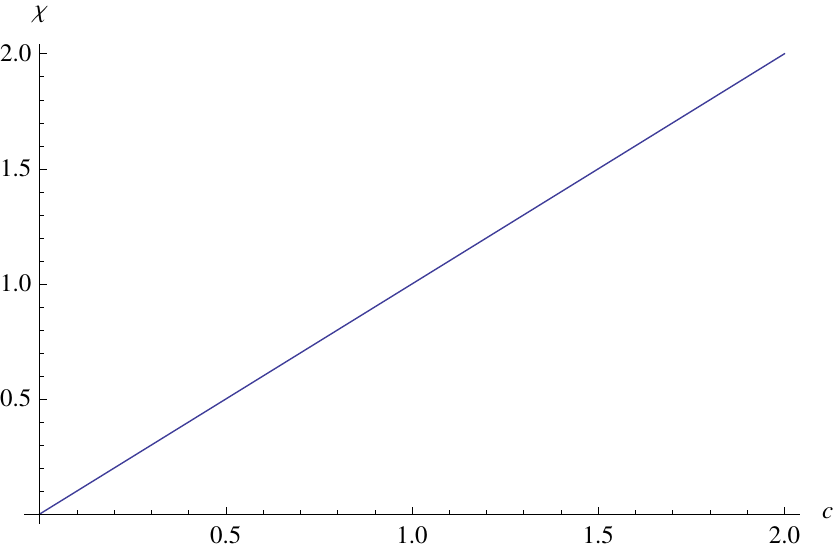} \caption{Information  recovered by Bob,
quantified  by the  Holevo  quantity  $\chi$,
  as a function of the  key information $c$ communicated by
  Charlie, in the noiseless case.}
\label{fig:holevononoise} %
\end{figure}

We wish to  parametrize the amount of key  information Charlie reveals
by means  of a single variable  $\psi$. To this end,  we assume that
the   ensemble  Alice  and   Bob  receive   from  Charlie,   given  by
Eq. (\ref{eq:bell}),  is a Werner  state of the  kind 
\begin{equation}
  \rho_{AB}^{(0,0)}(\psi)=a\Pi_{0,0}+b\sum_{j,k\ne00}\Pi_{j,k}, 
\label{eq:werner}
\end{equation}
where $\Pi_{j,k}$ is projector to  the Bell state $B_{j,k}$, $a=0.25 +
0.75\sin(\psi)$,     $b     =     0.25(1-\sin(\psi))$      and
$b\equiv\frac{1-a}{3}$.  The amount of information provided by Charlie
is  $2-H(a,b,b,b)$ bits,  where  $H(\cdots)$ is  Shannon entropy.   If
$c=2$   ($a=1$),   then   Bob   knows   Charlie   had   sent   out   a
$|\Phi^{+}\rangle$ state, and can work out Alice's encoded information
via   Eq.   (\ref{eq:belltrans}).   Similarly   other  
Werner   states   are  possible.   
The  maximum information  Bob  can
extract from  this ensemble  is the Holevo  quantity for  the ensemble
(\ref{eq:werner}). Fig. (\ref{fig:holevononoise}) shows how Bob's information
increases with key information in the noiseless case.

\section{Noise considerations}

In  Ref. \cite{puncudge},  the  authors studied  the  effect of  phase
damping noise on the $c=1$ bit  protocol. Here we consider noise to be
a squeezed generalized amplitude damping channel \cite{sban} acting on
Alice's qubit  transmission. More realistically, we  expect noise also
to affect Charlie's transmissions to Alice and Bob, but as these added
complications offer no new theoretical insight, we ignore them in this
work. The  action of a squeezed generalized  amplitude damping channel
is given by the set of Kraus operators

\begin{eqnarray}
\begin{array}{ll}
E_{0}\equiv\sqrt{p_{1}}\left[\begin{array}{ll}
\sqrt{1-\alpha(t)} & 0\\
0 & \sqrt{1-\beta(t)}\end{array}\right];~~~~ & E_{1}\equiv\sqrt{p_{1}}\left[\begin{array}{ll}
0 & \sqrt{\beta(t)}\\
\sqrt{\alpha(t)}e^{-i\phi(t)} & 0\end{array}\right];\\
E_{2}\equiv\sqrt{p_{2}}\left[\begin{array}{ll}
\sqrt{1-\mu(t)} & 0\\
0 & \sqrt{1-\nu(t)}\end{array}\right];~~~~ & E_{3}\equiv\sqrt{p_{2}}\left[\begin{array}{ll}
0 & \sqrt{\nu(t)}\\
\sqrt{\mu(t)}e^{-i\theta(t)} & 0\end{array}\right].\end{array}\label{eq:new}\end{eqnarray}

The details of the various  parameters appearing in the above equation
can  be  obtained  from   \cite{sban}.  
In particular, they depend upon the bath squeezing parameter $r$ and 
temperature $T$.
It  is  readily  checked  that
Eq.       (\ref{eq:new})        satisfies       the       completeness
condition                                              \begin{equation}
  \sum_{j=0}^{3}E_{j}^{\dag}E_{j}=\mathbb{I},\end{equation}
provided \begin{equation} p_{1}+p_{2}=1.\label{eq:p1p2}\end{equation}

\begin{figure}
\includegraphics{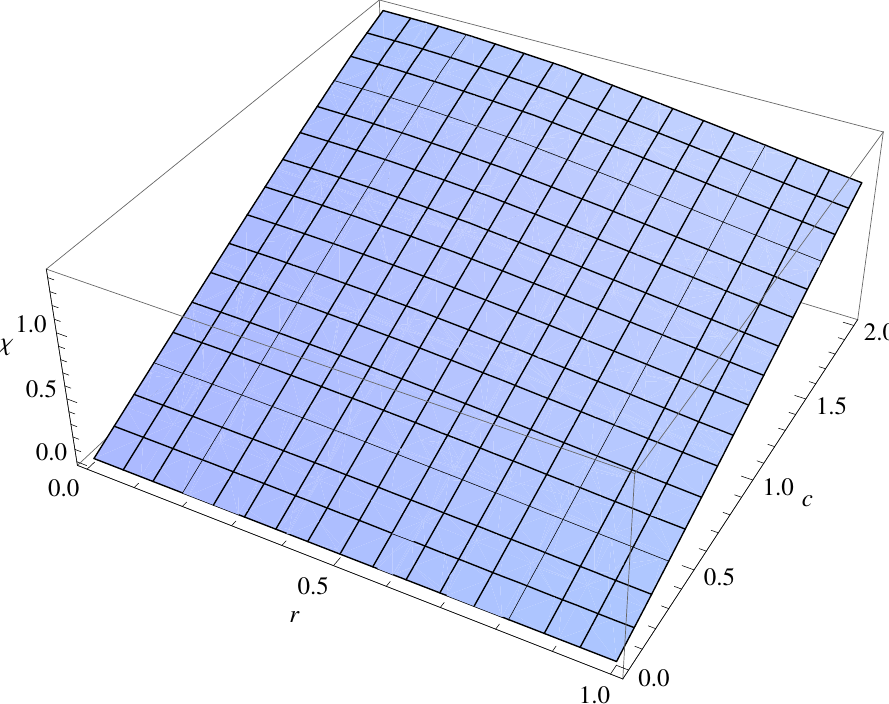} \caption{Information  recovered by Bob,
 quantified  by the  Holevo  quantity  $\chi$,
  as a function of the squeezing parameter $r$, coming from the Squeezed
Generalized Amplitude Damping Channel, and  key information $c$ 
communicated by Charlie. The time of evolution $t=0.5$, while temperature $T=0.1$.}
\label{fig:holevosq} %
\end{figure}

In  Fig.   (\ref{fig:holevosq}),  variation of Bob's recovered information, quantified by the Holevo
quantity  $\chi$, as a function of  bath squeezing $r$, and Charlie's  information $c$, is depicted.
The Holevo quantity $\chi$ increases with $c$, but not
as much as in  the noiseless case (Fig. (\ref{fig:holevononoise})): due to
the randomness introduced by the noise. Also, for a given information level $c$,
$\chi$ decreases with increase in squeezing $r$.

\begin{figure}
\includegraphics{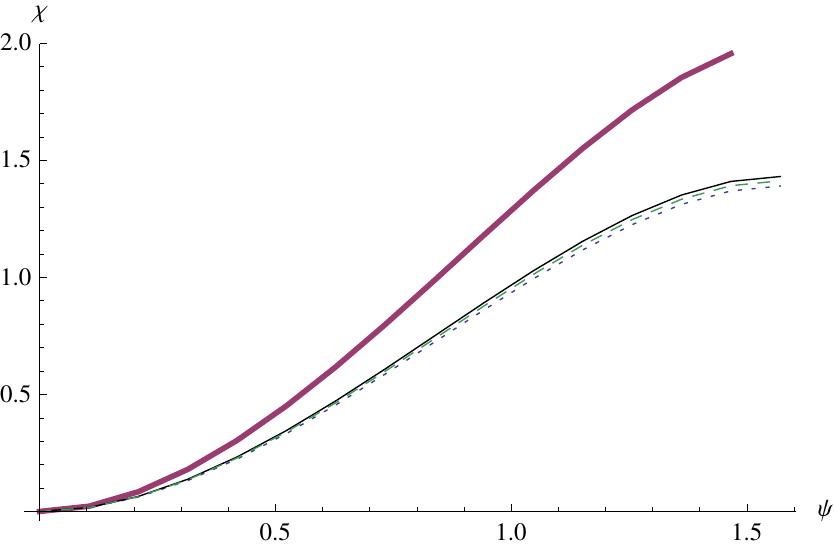} \caption{Information  recovered by Bob,
 quantified  by the  Holevo  quantity  $\chi$,
  as a function of the parameterizing angle $\psi$ (\ref{eq:werner}).
The thick bold line corresponds to the noiseless case, the thin bold line to the case
of noise with zero bath squeezing $r = 0$, while the large-dashed and small-dashed lines
correspond to noise with squeezing $r$ equal to -0.2 and 0.3,
respectively.  The time of evolution $t = 0.5$, while the temperature  
$T = 0.1$.}
\label{fig:holevo3} %
\end{figure}

The relationship between the Holevo quantity 
and the parameterizing angle $\psi$ (\ref{eq:werner}) is shown in  Fig. (\ref{fig:holevo3}). For all
cases shown the Holevo quantity $\chi$ increases with the angle $\psi$,
with $\psi = \pi/2$ corresponding to a pure Bell state.
Increase in angle $\psi$ corresponds to increase in the purity of the ensemble
(\ref{eq:werner}) leading to an increase in $\chi$. As expected, noise causes
a depletion in the Holevo quantity. The maximum difference in $\chi$, between 
the noiseless and noisy case, is seen to occur at $\psi = \pi/2$, that is,
when the state is pure the depletion due to noise is maximum.

\begin{figure}
\includegraphics{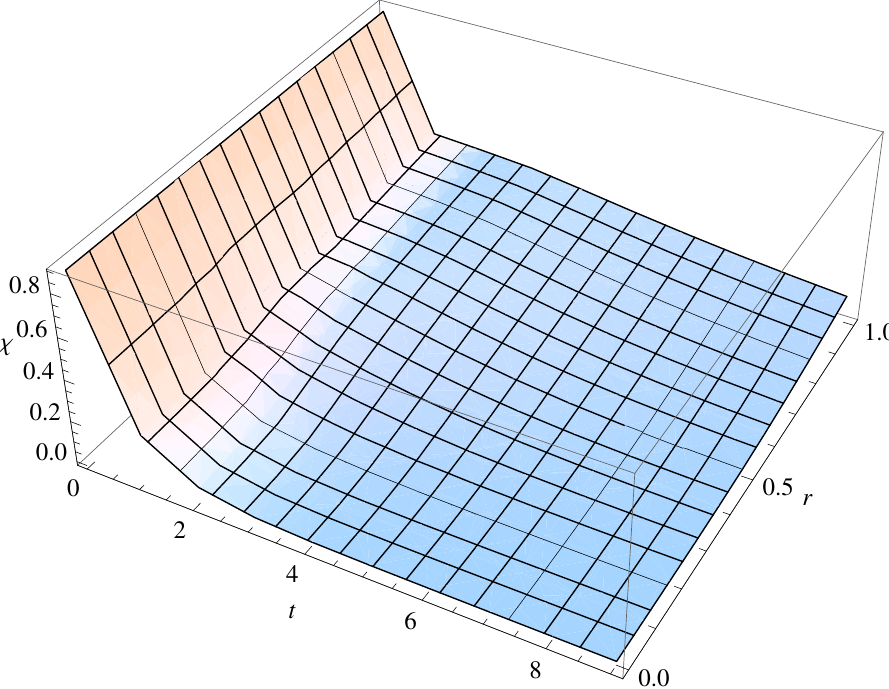}   \caption{Information   recovered  by   Bob,
   quantified  by the  Holevo  quantity  $\chi$, as  a  function of  the
   squeezed generalized amplitude  damping channel parameters $r$ 
	(squeezing) and
   $t$ (time of evolution), assuming
   Charlie communicates one bit of information.
   We note that, for sufficiently early times, squeezing fights thermal
	effects (here $T=0.1$) to cause an increase in the recovered
	information. }  
\label{fig:holevochr} %
\end{figure}

Fig. (\ref{fig:holevochr}) shows the effect, on the Holevo quantity, 
of squeezing as a function
of time, in particular, demonstrating its favorable influence in
certain regimes. As seen from the figure, for sufficiently early times, 
squeezing fights thermal effects to cause an increase in the 
recovered information.

% The
% dependence  of Bob's  information, with  $c=2$, and  as a  function of
% channel   parameters   $\lambda$   and    $p$   is   shown   in   Fig.
% (\ref{fig:3DPlot}).  
The  significance of  noise  is that
Alice and  Bob may  consume some  of the Bell  pairs to  determine the
noise  level, and  decide  whether it  is  too high  to permit  secure
information  transfer.  

\section{Cryptographic switch without any restriction on the channel}

Let us withdraw the restriction imposed on the channel from Bob to
Alice. Now we may modify the proposed protocol as follows to remain
secured:
\begin{enumerate}
\item  After  receiving Alice's  request,  Charlie  prepares $n$  Bell
  states (not  all the same).  He uses the  Bell states to  prepare an
  ordered    sequence    with     all    the    first    qubits    as,
  $P_{A}=\left[p_{1}\left(t_{A}\right),p_{2}\left(t_{A}\right),...,
p_{n}\left(t_{A}\right)\right]$,
  and               another              ordered              sequence
  $P_{B}=[p_{1}(t_{B}),p_{2}(t_{B}),...,p_{n}(t_{B})]$  with  all  the
  second  qubits resulting in  an ordered  sequence where  the subscript
  $1,2,...,n$    denotes    the    order    of   a    particle    pair
  $p_{i}=\{t_{A}^{i},t_{B}^{i}\},$ which is in the Bell state.
\item Charlie  scrambles the second  qubits: that is, he  disturbs the
  order  of   the  qubits  in   $P_{B}$  to  create  a   new  sequence
  $P_{B}^{\prime}=[p_{1}^{\prime}(t_{B}),p_{2}^{\prime}(t_{B}),...,p_{n}^{\prime}(t_{B})]$
  and sends  it to  Bob. The  actual order is  known to  Charlie only.
  This ordering of information is part of the \textit{extended key} of
  the modified protocol.
\item After receiving the  qubits from Charlie, Alice understands that
  she has  been permitted to send  the information to  Bob.  Since the
  sequence  with Alice and  Bob are  different, even  if Alice  or Bob
  obtain the  access of both  $P_{A}$ and $P_{B}^{\prime}$,  they will
  not be  able to find out  the Bell states prepared  by Charlie. Thus
  any kind of collusion between Alice and Bob would fail.
\item Alice uses dense coding to encode two bits of classical information
on each qubit and transmits her qubits to Bob.
\item When Charlie  plans to Allow Bob to  know the secret information
  communicated to Bob by Alice, then Charlie discloses the Bell states
  which he had prepared and the exact sequence.
\item Since the  initial Bell states and exact  sequence is known, Bob
  now measures  his qubits in  Bell basis and obtains  the information
  encoded by Alice.
\end{enumerate}

\section{Discussion and conclusions}

The proposed protocol allows to secretly share classical information
in presence of a noisy quantum channel. In the protocol Alice shares
the secret classical information with Bob but a control is kept with
Charlie. Thus the role of Charlie may be viewed as that of a switch,
which controls the channel between Alice and Bob. The protocol may
also be viewed as a protocol of controlled secret sharing or a protocol
of controlled dense coding. In the sequences $P_{A}$ and $P_{B}^{\prime}$
Charlie can also insert some decoy photons and use them as check bits
to ensure that there is no eavesdropping during his communication
with Alice and Bob.

In Section II we assumed the  channel as one-way and in Section III we
considered the effect  of noise.  The one-way channel  assumption is a
reasonable assumption  when the channel  is noisy. This is  so because
the  rate of  success of  an  effort in  which they  illegally try  to
ascertain the Bell state sent by Charlie, for example by Alice sending
her qubit  to Bob first and  Bob measuring the Bell  state prepared by
Charlie and re-sending the  qubit to Alice for dense-coding operation,
will be very  small. This is so because it  would require particles of
the Bell state to travel thrice through the noisy channel.  High noise
level opens the  danger of an eavesdropper that  neither Alice nor Bob
would  desire.  This  potentially  provides an  instance  where  noisy
channel communication is useful.

The restriction  on the channel is  removed in Section IV  and we have
seen  that our  modified protocol  works in  a very  general scenario.
Since it is easy to  experimentally generate Bell states, the proposed
cryptographic switch  may be realized  experimentally and can  be used
for  many  practical problems  of  similar  kind,  say, in  which  the
director of an organization wishes to keep a control over the time and
amount  of  information  to  be   disclosed  to  an  employee  of  the
company. The  custodian of  the files (Alice)  must not be  worse than
semi-honest, as  she could otherwise create  her own classical/quantum
channel  and communicate  directly with  Bob.  But  there is  always a
potential chance that Charlie can detect such communication.  Further,
it is reasonable  to assume that as a  subordinate in an organization,
she  lacks  the  resources  to  create or  purchase  the  entanglement
required.  Further,  the protocol is  also valid even when  Charlie is
semi-honest, fully honest Charlie being  one who does not get any part
of the information sent by Alice to Bob.

We have visualized the problem  of cryptographic switch as a practical
problem  in a  particular  situation. Many  analogous examples  exist,
where similar situations arise. For  example, one may think that the owner
of a company (Charlie) has  asked his semi-honest assistant (Alice) to
send details  of all his  shares to a  stock exchange broker,  Bob, to
sell it  in the  stock market. But  Charlie wishes  to keep an  eye on
stock  fluctuations and  to  permit Bob  to  sell his  shares at  some
suitable  time. These  are only  some  specific examples,  the idea  of
cryptographic  switch is  expected  to be  useful  in various  analogous
situations  of  practical relevance,  the  idea  being  that the  main
information  is  priorly sent  using  a  secure  quantum channel,  and
`switched on' by a small classical key, when convenient.

%\begin{acknowledgments}

\textbf{Acknowledgments:} The work of A. P. is supported by Department of Science and Technology, India under the project no. SR/S2/LOP/0012/2010.

\end{document}